\documentclass[11pt,xcolor=dvipsnames, reqno]{amsart}
\usepackage[a4paper, top=3cm, bottom=4cm, left=3cm, right=3cm]{geometry}
\usepackage[T2A, T1]{fontenc}
\usepackage[utf8]{inputenc}

\usepackage[english]{babel}
\usepackage{amsmath,amssymb,amstext}
\usepackage{amsthm}
\usepackage{amsfonts}
\usepackage{amsxtra}
\usepackage{graphicx}
\usepackage{wrapfig}
\usepackage[dvipsnames]{xcolor}
\usepackage{subcaption}
\usepackage{braket}
\usepackage{enumerate}
\usepackage[super]{nth}
\usepackage{pdfpages}
\usepackage{graphicx}
\usepackage{svg}
\usepackage{tikz}

\usepackage{hyperref}
\usepackage[backend=bibtex,style=numeric,sorting=none,firstinits=true, maxbibnames=99]{biblatex}

\usepackage{xurl}
\hypersetup{breaklinks=true}

\usepackage{xifthen}

\allowdisplaybreaks

\bibliography{Literature.bib}

\newcommand{\R}{\mathbb{R}}
\newcommand{\N}{\mathbb{N}}

\newcommand{\cH}{\mathcal{H}}

\newcommand{\dr}[1]{\mathrm{d}{#1}\, }

\newcommand{\ga}{\gamma}
\newcommand{\be}{\beta}

\newcommand{\pt}{\tilde p}

\newcommand{\om}{\Omega}

\newcommand\no[2]{\left\lVert#1\right\rVert_{#2}}

\newcommand\ev[1]{N\left(-\Delta_{#1}^{N}+V\right)}
\newcommand\evv[2]{N\left(-\Delta_{#1}^{N}+#2\right)}

\newcommand\evd[1]{N\left(-\Delta_{#1}^{D}+V\right)}
\newcommand\inx[2]{\int_{#1}\dr{#2}}

\let\epsilon\varepsilon
\let\phi\varphi

\makeatletter
\providecommand*{\cupdot}{%
  \mathbin{%
    \mathpalette\@cupdot{}%
  }%
}
\newcommand*{\@cupdot}[2]{%
  \ooalign{%
    $\m@th#1\cup$\cr
    \sbox0{$#1\cup$}%
    \dimen@=\ht0 %
    \sbox0{$\m@th#1\cdot$}%
    \advance\dimen@ by -\ht0 %
    \dimen@=.5\dimen@
    \hidewidth\raise\dimen@\box0\hidewidth
  }%
}

\providecommand*{\bigcupdot}{%
  \mathop{%
    \vphantom{\bigcup}%
    \mathpalette\@bigcupdot{}%
  }%
}
\newcommand*{\@bigcupdot}[2]{%
  \ooalign{%
    $\m@th#1\bigcup$\cr
    \sbox0{$#1\bigcup$}%
    \dimen@=\ht0 %
    \advance\dimen@ by -\dp0 %
    \sbox0{\scalebox{2}{$\m@th#1\cdot$}}%
    \advance\dimen@ by -\ht0 %
    \dimen@=.5\dimen@
    \hidewidth\raise\dimen@\box0\hidewidth
  }%
}
\makeatother

\newcommand{\normiii}[1]{{\left\vert\kern-0.25ex\left\vert\kern-0.25ex\left\vert #1 
    \right\vert\kern-0.25ex\right\vert\kern-0.25ex\right\vert}}

\newtheorem{theorem}{Theorem}[section]

\allowdisplaybreaks

\title[Weyl's law for Neumann Schr\"odinger operators on H\"older domains]{Weyl's law for Neumann Schr\"odinger operators on H\"older domains}
\author{Charlotte Dietze}
\address{Department of Mathematics, LMU Munich, Theresienstr.~39, 80333 Munich, Germany\newline
Institut des Hautes Études Scientifiques, 35 route de Chartres, 91440 Bures-sur-Yvette, France}
\email{dietze@math.lmu.de}
\date{\today}

\makeatletter
\@namedef{subjclassname@2020}{\textup{2020} Mathematics Subject Classification}
\makeatother

\subjclass[2020]{35P15, 35P20}
\keywords{Neumann Laplacian, H\"older domains, Cwikel-Lieb-Rozenblum inequality, Semiclassical asymptotics, Weyl's law.}

\begin{document}
\maketitle
\begin{abstract}
We review recent results on the semiclassical behaviour of Schr\"odinger operators with Neumann boundary conditions. In this setting, the validity of Weyl's law requires additional conditions on the potential. We will explain the techniques needed to control the number of bound states near the boundary, thus leading to universal estimates on the number of bound states.
\end{abstract}

\tableofcontents

\section{Introduction}
Weyl's law for the eigenvalues of the Laplacian on a domain $\Omega\subset\R^d$ (a bounded, open and connected subset of $\R^d$) states that the number of eigenvalues below $\lambda$, which we denote by $N\left(- \Delta_\Omega - \lambda\right)$, satisfies 
\begin{equation}\label{eq:intweyldconst}
    N\left(- \Delta_\Omega - \lambda\right) = \frac{|B_1^d(0)|}{(2\pi)^d} |\Omega|\lambda^\frac{d}{2} + o\left(\lambda^\frac{d}{2}\right) \mathrm{\ as\ } \lambda \rightarrow \infty,
\end{equation}
under suitable assumptions on the boundary conditions or the domain $\Omega$, where $|B_1^d(0)|$ is the volume of the unit ball in $\R^d$.

\bigskip

Its history, see \cite{arendt2009weyl}, goes back to Rayleigh \cite{rayleigh1896theory} in 1877 who examined the number of overtones of musical instruments such as a violin string or an organ pipe. In three dimensions, he derived that for a cubic organ pipe, the number of overtones, which are the square root of the eigenvalues of the Laplacian, below the frequency $\nu$ behaves like the volume of the organ pipe multiplied by $\nu^3$ for large $\nu$. He could later connect this to the famous blackbody radiation experiments by Planck in the 1890s \cite{planck1901ueber}, where he could derive the correct behaviour of the emitted energy for cubical shapes \cite{rayleigh1905dynamical}. Sommerfeld \cite{sommerfeld1910greensche} and Lorentz \cite{lorentz1910alte} remarked in 1910 that it remains to show that this law for the eigenvalues of the Laplacian does not depend on the shape considered. This problem was solved in 1911 by Weyl \cite{weyl1911asymptotische}, thereby rigourously justifying \eqref{eq:intweyldconst} for the Dirichlet Laplacian $- \Delta_\Omega^D$\footnote{The Dirichlet Laplacian is the self-adjoint operator on $L^2(\om)$ corresponding to the quadratic form $\int_\om  |\nabla u|^2$ defined on $H^1_0(\om)$}. For his proof, Weyl further developed the min-max principle, which was introduced by Fischer \cite{fischer1905quadratische} and also later by Courant \cite{courant1920eigenwerte}. Weyl showed \eqref{eq:intweyldconst} for bounded domains $\Omega\subset\R^d$ with sufficiently smooth boundary. Rozenblum extended \eqref{eq:intweyldconst} to  all open sets $\Omega\subset\R^d$ of finite measure \cite{rozenblum1971distribution, rozenbljum1972eigenvalues}.

\bigskip

\begin{figure}[htbp]
  \centering 
  \includegraphics{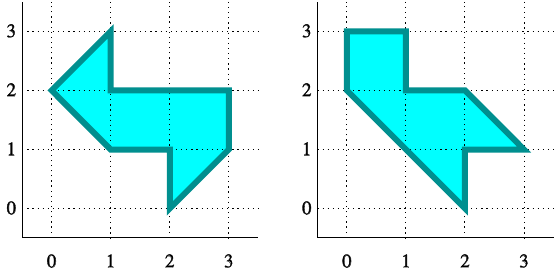} 
  \caption{This image taken from \cite{drumwiki} shows two examples of ``drums that sound the same'' in the sense that the Dirichlet Laplacian on both domains has the same spectrum, see \cite[Figure 5]{buser2010some} and \cite[Figure 10]{gordon1992isospectral}. } 
  \label{fi:drum}
\end{figure}

The eigenvalue problem of the Laplacian was also made popular by Kac in his 1966 paper ``Can one hear the shape of a drum?'' \cite{kac1966can}. The general answer is no \cite{milnor1964eigenvalues, gordon1992isospectral, buser2010some}, see Figure \ref{fi:drum}, except if $\Omega$ is analytic and satisfies certain symmetry properties \cite{zelditch2009inverse}. Nevertheless, in any case, we can always ``hear'' the volume $|\Omega|$ from Weyl's law \eqref{eq:intweyldconst}. Also, due to a classical result of Ivrii \cite{ivrii1980second}, if $\Omega$ is sufficiently smooth, then we can also ``hear'' the surface area $ |\partial \Omega|$ from a second-order Weyl law 

\begin{equation}\label{eq:intweyldconstivrii}
    N\left(- \Delta_\Omega^D - \lambda\right) = \frac{|B_1^d(0)|}{(2\pi)^d} |\Omega|\lambda^\frac{d}{2} -\frac{1}{4}\frac{|B_1^{d-1}(0)|}{(2\pi)^{d-1}}|\partial \Omega|\lambda^\frac{d-1}{2}
    + o\left(\lambda^\frac{d-1}{2}\right) \mathrm{\ as\ } \lambda \rightarrow \infty.
\end{equation}
See also \cite{Frank-Larson} for an extension of \eqref{eq:intweyldconstivrii} to the sum of eigenvalues which holds for all Lipschitz domains. Here $\Omega$ is called a Lipschitz domain if it is a domain and the boundary of $\Omega$ is locally the graph of a Lipschitz continuous function $f$, that is, there exists a constant $c>0$ such that $|f(x)-f(y)|\le c|x-y|$ for all $x,y$ in the domain of $f$.

\bigskip

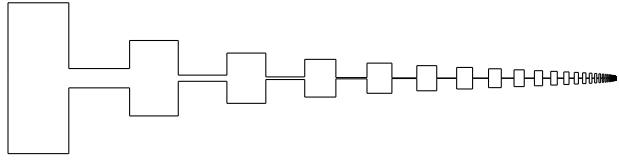
\begin{figure}[htbp]
  \centering 
  \begin{tikzpicture}
\draw (0,0) -- ++(0,1);
\foreach \i in {1,...,159}{
		\draw ({2*(5*(1-(4/5)^(\i))-1)},{1/(\i)}) -- ++({(4/5)^(\i)},0);
}
\foreach \i in {1,...,159}{
		\draw ({2*(5*(1-(4/5)^(\i))-1)+(4/5)^(\i)},{1/(5*\i)*(4/5)^(2*\i)}) -- ++({(4/5)^(\i)},0);
}
\foreach \i in {1,...,159}{
		\draw  ({2*(5*(1-(4/5)^(\i))-1)+(4/5)^(\i)},{1/(\i)})--({2*(5*(1-(4/5)^(\i))-1)+(4/5)^(\i)},{1/(5*\i)*(4/5)^(2*\i)});
}

\foreach \i in {1,...,159}{
		\draw  ({2*(5*(1-(4/5)^(\i))-1)+2*(4/5)^(\i)},{1/(5*\i)*(4/5)^(2*\i)})--({2*(5*(1-(4/5)^(\i))-1)+2*(4/5)^(\i)},{1/(\i+1)});
}

\draw (0,0) -- ++(0,-1);
\foreach \i in {1,...,159}{
		\draw ({2*(5*(1-(4/5)^(\i))-1)},{-1/(\i)}) -- ++({(4/5)^(\i)},0);
}
\foreach \i in {1,...,159}{
		\draw ({2*(5*(1-(4/5)^(\i))-1)+(4/5)^(\i)},{-1/(5*\i)*(4/5)^(2*\i)}) -- ++({(4/5)^(\i)},0);
}
\foreach \i in {1,...,159}{
		\draw  ({2*(5*(1-(4/5)^(\i))-1)+(4/5)^(\i)},{-1/(\i)})--({2*(5*(1-(4/5)^(\i))-1)+(4/5)^(\i)},{-1/(5*\i)*(4/5)^(2*\i)});
}

\foreach \i in {1,...,159}{
		\draw  ({2*(5*(1-(4/5)^(\i))-1)+2*(4/5)^(\i)},{-1/(5*\i)*(4/5)^(2*\i)})--({2*(5*(1-(4/5)^(\i))-1)+2*(4/5)^(\i)},{-1/(\i+1)});
}


\end{tikzpicture}
  \caption{This image shows an example of a domain referred to as ``rooms and passages'', which goes back to  Hempel and Seco \cite{hempel1991essential}. For such a domain, zero is contained in the essential spectrum of the Neumann Laplacian and in particular, Weyl's law does \emph{not} hold.} 
  \label{fi:roomspassages}
\end{figure}

The results we presented above were stated for the Dirichlet Laplacian $\Delta_\Omega^D$. Since we can trivially extend functions in that by zero outside $\Omega$, we can think of results for the Dirichlet Laplacian as ``global'' properties of the Laplacian. On the other hand, the Neumann Laplacian $\Delta_\Omega^N$\footnote{The Neumann Laplacian is the self-adjoint operator on $L^2(\om)$ corresponding to the quadratic form $\int_\om  |\nabla u|^2$ defined on $H^1(\om)$} appears naturally when localising. In order to understand its properties, one needs to better understand the geometry of the boundary since functions in $H^1(\om)$ can grow to infinity close to the boundary of $\om$. This makes many problems for the Neumann Laplacian more difficult than their analogues for the Dirichlet Laplacian. While many of the results mentioned above have a corresponding counterpart for the Neumann Laplacian under suitable assumptions, even very basic properties can fail for the Neumann Laplacian in general. While the Dirichlet Laplacian on a domain (which we always assume to be bounded) always has compact resolvent, there are examples of domains such that zero is in the essential spectrum of the Neumann Laplacian on that domain. For instance, Hempel and Seco \cite{hempel1991essential} constructed such a domain known as ``rooms and passages'', see Figure \ref{fi:roomspassages}. 

\bigskip

In a remarkable work, Netrusov and Safarov showed  Weyl's law for $\gamma$-Hölder domains with Neumann boundary conditions
\begin{equation}\label{eq:nsintweylnconst}
    N\left(- \Delta_\Omega^N - \lambda\right) = \frac{|B_1^d(0)|}{(2\pi)^d} |\Omega|\lambda^\frac{d}{2} + o\left(\lambda^\frac{d}{2}\right) \mathrm{\ as\ } \lambda \rightarrow \infty,
\end{equation}
holds for all $\ga\in\left(\frac{d-1}{d},1\right)$ \cite[Corollary 1.6]{netrusov2005weyl}, and it fails for all $\ga\in\left(0,\frac{d-1}{d}\right]$ in the sense that for those $\ga$, there exists a $\ga$-Hölder domain  $\om$ such that \eqref{eq:nsintweylnconst} is not true \cite[Theorem 1.10]{netrusov2005weyl}. Here $\Omega$ is called a $\ga$-Hölder domain if it is a domain and the boundary of $\Omega$ is locally the graph of a $\ga$-Hölder continuous function $f$, that is,  there exists a constant $c>0$ such that $|f(x)-f(y)|\le c|x-y|^\gamma$ for all $x,y$ in the domain of $f$. Note that the case $\ga=1$ corresponds to Lipschitz domains, which are well known to satisfy \eqref{eq:nsintweylnconst}, see \cite[Theorem 3.20]{frank2021schrodinger}.

\bigskip

The reason why there is a transition at $\ga=\frac{d-1}{d}$  can be intuitively explained as follows. Locally, the boundary of $\Omega$ is the graph of a Hölder continuous function
\begin{equation}\label{eq:fA}
    f:A\to\R\quad \text{with}\quad A\subset\R^{d-1}\quad \text{open and bounded}, 
\end{equation}
that is, it is a subset of the boundary of $\partial\om$ up to a local change of coordinates by translation and rotation. Let us denote by $\cH^s$ is the $s$-dimensional Hausdorff measure for $s>0$. Then there exists a constant $C>0$ such that 
\begin{equation}\label{eq:hausd}
    \cH^{\frac{d-1}{\ga}}\left(\left\{(x',f(x'))\ \mid \, x'\in A\right\}\right)\le C\cH^{d-1}\left(A\right)<\infty.
\end{equation}
It follows that the Hausdorff dimension of $\partial\om$ is at most $\frac{d-1}{\ga}$. If $\ga>\frac{d-1}{d}$, then by \eqref{eq:hausd}, the Hausdorff dimension of $\left\{(x',f(x'))\ \mid \, x'\in A\right\}$  is strictly smaller than $d$, which is the Hausdorff dimension of $\om$. Hence, the contribution from the  bulk of $\om$ should dominate the boundary effects. If $\ga\le \frac{d-1}{d}$, then the relatively straightforward proof of \eqref{eq:hausd} is not enough to decide if the Hausdorff dimension of the boundary $\partial\om$ is equal to $d$ (note that it cannot be larger since $\partial\om\subset\R^d$). The boundary effects might be of the same or even higher order as the contribution from the bulk of $\om$ and this is indeed what one observes for Weyl's law.

\bigskip

The proof idea by Netrusov and Safarov for \eqref{eq:nsintweylnconst} is to decompose the domain $\om$ into smaller domains on which there is at most one negative eigenvalue of $- \Delta^N - \lambda$ each. Then the number of oscillatory domains chosen gives an upper bound for the number of negative eigenvalues $N\left(- \Delta_\Omega^N - \lambda\right)$ of $- \Delta_\Omega^N - \lambda$.

\bigskip

It will turn out that for $\ga>\frac{d-1}{d}$, the parts close to the boundary only give a subleading contribution, so the leading order contribution comes from the oscillatory domains in the bulk of $\om$. This leading order contribution can be shown to be the right-hand side of \eqref{eq:nsintweylnconst} in the same way as in Weyl's proof for Weyl's law, thereby establishing the upper bound. The proof of the lower bound is simpler, for example by comparing with the Dirichlet Laplacian.

\bigskip

Another way to view Weyl's law \eqref{eq:intweyldconst} is via semiclassics. It suggests that every bound state corresponds to a volume of size $(2\pi)^d$ in the phase space \cite[Section 4.1.1]{lieb2010stability}, which is in our case given by $\om \times \R^d$. For large $\lambda$, the semiclassical approximation states
\begin{align}\label{eq:semiconst}
    N\left(- \Delta_{\Omega} - \lambda\right)  \approx \frac{1}{(2\pi)^{d} } \left|\left\{(p,x) \in \R^d \times \Omega \, \bigm| \,  |p|^2 - \lambda <0 \right\}\right| =  \frac{|B_1^d(0)|}{(2\pi)^d} |\Omega|\lambda^\frac{d}{2} ,
\end{align}
compare with \eqref{eq:intweyldconst}.

\bigskip

In the following, we will focus on Schr\"odinger operators and we will present the results from \cite{dietze2023semiclassical}. To this end, let  $V: \Omega \to (-\infty,0]$ be measurable. We will refer to $V$ as a potential. By the semiclassical approximation, we expect for a large $\lambda$
\begin{align}\label{eq:semipot}
    N\left(- \Delta_{\Omega} + \lambda V \right)  \approx \frac{1}{(2\pi)^{d} } \left|\left\{(p,x) \in \R^d \times \Omega \, \bigm| \,  |p|^2 + \lambda V(x) <0 \right\}\right| =  \frac{|B_1^d(0)|}{(2\pi)^d} \lambda^\frac{d}{2} \int_{\Omega} |V|^{\frac d 2}.
\end{align}

While Netrusov and Safarov showed that Weyl's law holds for $\ga$-Hölder domains with $\ga\in\left(\frac{d-1}{d},1\right)$, our first result  \cite[Theorem 1.1]{dietze2023semiclassical} shows that the situation for  Schr\"odinger operators is more delicate.

\begin{theorem}[Example with non-semiclassical behaviour]\label{th:example} Let $d\ge 2$.     For every $\gamma \in \left(\tfrac{d-1}{d}, 1\right)$ there exists a $\gamma$-Hölder domain $\Omega \subset \R^d$ and $V : \Omega \rightarrow \left( -\infty, 0\right]$ with $V \in L^\frac{d}{2}(\Omega)$ such that
    \begin{equation}\label{eq:thexeq}
        \limsup_{\lambda \rightarrow \infty} \frac{\evv{\Omega}{\lambda V}}{\lambda^\frac{d}{2}} = \infty .
    \end{equation}
\end{theorem}

Theorem \ref{th:example} shows that in general, we cannot even expect semiclassical behaviour, and in particular, the semiclassical approximation does not hold in the setting of Theorem \ref{th:example}. However, if we make further assumptions on the potential $V$, we can prove a universal Cwikel-Lieb-Rozenblum-type bound on the number of negative eigenvalues of the Schr\"odinger operator $- \Delta_{\Omega}^N +  V$, see  \cite[Theorem 1.2]{dietze2023semiclassical}.

\begin{theorem}[Cwikel-Lieb-Rozenblum type bound]\label{th:weightednorm}
   Let $d\ge 2$. Let $\gamma \in \left[\tfrac{2(d-1)}{2d - 1} , 1 \right)$ and let $\Omega$ be a $\gamma$-Hölder domain. Then there exists a constant $C_\om = C_\om(d, \gamma, \Omega)> 0$ and $p_{d,\gamma}>\frac d 2 $ such that for every $V : \Omega \rightarrow \left(- \infty, 0 \right]$ with $V\in L^{p_{d,\gamma}}(\om)$
       \begin{equation}\label{eq:clrthm}
        \ev{\Omega} \leq C_\om\left(1 + \no{V}{p_{d,\gamma}}^\frac{d}{2}\right) .
    \end{equation}
   Moreover,  $p_{d,\gamma} $ satisfies
       \begin{equation}
   \lim_{\gamma\to 1}     p_{d,\gamma} = \frac{d}{2}. 
     \end{equation}
\end{theorem}


The main point of Theorem \ref{th:weightednorm} is that we obtain the expected semiclassical behaviour 
\begin{equation}\label{eq:clrnicesemib}
    \evv{\om}{\lambda V}=\mathcal{O}\left(\lambda^{\frac{d}{2}}\right) \ \textrm{as } \lambda\to\infty
\end{equation}
if $V\in L^{p_{d,\gamma}}(\om)$. It is possible to obtain for all $\ga\in\left(0,1\right)$ 
\begin{equation}\label{eq:pclr}
        \ev{\Omega} \leq C_\om\left(1 + \int_\om |V|^{\pt}\right) .
    \end{equation}
for some $\pt>\frac{d}{2}$ by following the strategy of Rozenblum, see below in \eqref{eq:clrbad} and see also \cite{frank2010equivalence}  combined with \cite{labutin}. However, this bound is insufficient for \eqref{eq:clrnicesemib}. 

\bigskip

While  Theorem \ref{th:weightednorm} is of independent interest, we can use it to derive Weyl's law on Hölder domains, thus rigourously justifying \eqref{eq:semipot}, see \cite[Theorem 1.3]{dietze2023semiclassical}.


\begin{theorem}[Weyl's law for Schr\"odinger operators on H\"older domains]\label{th:weyllawforapotential} Let $d\ge 2$. Let $\gamma \in \left[\tfrac{2(d-1)}{2d - 1} , 1 \right)$ and let $\Omega \subset \R^d$ be a $\gamma$-Hölder domain. Let $V : \Omega \rightarrow (-\infty, 0]$ with $V\in L^{p_{d,\gamma}}(\om)$. Then
\begin{equation}\label{eq:weylthm}
    N \left(-\Delta^N_\Omega + \lambda V\right) = (2 \pi)^{-d} \left|B_1^d(0)\right| \lambda^{\frac{d}{2}} \int_\Omega |V|^{\frac{d}{2}} + o \left(\lambda^{\frac{d}{2}}\right) \mathrm{\ as\ } \lambda \rightarrow \infty .
\end{equation}
\end{theorem}
The proof idea for Theorem \ref{th:weyllawforapotential} is to approximate the potential $V$ by a sequence of potentials $\left(V_n\right)_{n\in\N}$, which are continuous and compactly supported inside $\om$. For each of the potentials $V_n$, we can follow Weyl's proof strategy for the Weyl law for constant potentials to obtain Weyl's law for the Schr\"odinger operator $-\Delta^N_\Omega + \lambda V_n$. For instance, for the upper bound in \eqref{eq:weylthm}, we split for $\delta\in(0,1)$ 
\begin{equation}\label{eq:weylvvncomp-intro}
    N\left(-\Delta^N_\Omega + \lambda V \right) \leq N\left((1 - \delta) \left(-\Delta^N_\Omega\right) + \lambda V_n\right) + N \left(\delta \left(-\Delta^N_\Omega\right) + \lambda \left(V - V_n\right)\right) ,
   \end{equation}
and we divide by $\lambda^{\frac{d}{2}}$ and let $\lambda\to\infty$ first, then $n\to\infty$ and finally $\delta\to0$. We will recover the right-hand side in \eqref{eq:weylthm} from the first summand on the right-hand side in \eqref{eq:weylvvncomp-intro}. We will show that the contribution from  the second summand $N \left(\delta \left(-\Delta^N_\Omega\right) + \lambda \left(V - V_n\right)\right)$ goes to zero using Theorem \ref{th:weightednorm}.

\bigskip

The range of $\gamma$ in Theorem \ref{th:weyllawforapotential} is smaller than the optimal range $\gamma \in \left(\frac{d-1}{d}, 1\right)$ obtained in  \cite{netrusov2005weyl} for constant potentials. In fact, we are able to cover the full optimal range including the endpoint $\gamma \in \left[\frac{d-1}{d}, 1\right)$ provided we replace the $L^{p_{d,\gamma}}(\om)$ norm by a weighted $L^{\pt}$ norm $\normiii{V}$ with $\pt=\pt(d,\ga)>\frac{d}{2}$, which gives better control on the growth of the potential near the boundary, see \cite[Theorem 1.3]{dietze2023semiclassical}. The proof relies on an improved version of Theorem \ref{th:weightednorm} for $\gamma \in \left[\frac{d-1}{d}, 1\right)$ if $\normiii{V}<\infty$, see Theorem \ref{th:weightednormfullrange} below and also \cite[Theorem 1.2]{dietze2023semiclassical}. The norm $\normiii{V}$ controls the growth of the potential $V$ near the boundary $\partial\om$. Note that the domain $\om$ and the potential in Theorem \ref{th:example} do not satisfy \eqref{eq:semipot}, so $\normiii{V}=\infty$ in Theorem \ref{th:example}. Indeed, our construction in the proof of Theorem \ref{th:example} involves a potential $V$ that grows to infinity near the boundary of $\om$. 

\bigskip

We hope that the techniques we developed to deal with rough domains will be helpful for the investigation of semiclassics of potentials that are singular or oscillatory along some lower-dimensional manifold.

\bigskip

{\bf Acknowledgements.}  The author would like to express her deepest gratitude to Phan Thành Nam for his continued support and very helpful advice. She would also like to thank Laure Saint-Raymond for her support and hospitality at Institut des Hautes Études Scientifiques and for inspiring discussions. The author acknowledges the support from the Deutsche Forschungsgemeinschaft (DFG project Nr.~426365943), from the Jean-Paul Gimon Fund and from the Erasmus+ programme.

\section{The case of constant potentials}\label{s:ns}
In this section, we will explain the proof strategy of Netrusov and Safarov for Weyl's law for constant potentials \eqref{eq:nsintweylnconst} \cite[Corollary 1.6]{netrusov2005weyl}, see the following Theorem.
\begin{theorem}[Weyl's law for for constant potentials on H\"older domains]\label{th:weylns}
    Let $\gamma \in \left(\tfrac{d-1}{d} , 1 \right)$ and let $\Omega \subset \R^d$ be a $\gamma$-Hölder domain.  Then
\begin{equation}\label{eq:weylnsthm}
    N \left(-\Delta^N_\Omega - \lambda \right) = (2 \pi)^{-d} \left|B_1^d(0)\right| \left|\om\right| \lambda^{\frac{d}{2}}  + o \left(\lambda^{\frac{d}{2}}\right) \mathrm{\ as\ } \lambda \rightarrow \infty .
\end{equation}
\end{theorem}


    
The main idea is to cover the domain $\om$ by smaller domains, which we will call oscillatory domains in the following, such that there is at most one negative eigenvalue of the corresponding Schr\"odinger operator on each oscillatory domain. The number of negative eigenvalues on the domain can then be bounded by the number of oscillatory domains we have chosen.

\bigskip

Netrusov and Safarov can choose the oscillatory domains in the bulk of $\Omega$ as cubes of a fixed $\lambda$-dependent side length which are arranged on a lattice, and which only overlap on their boundaries.

\bigskip






Close to the boundary $\partial\om$ Netrusov and Safarov \cite{netrusov2005weyl} construct  the oscillatory domains as follows. They use the $(d-1)$-dimensional Besicovitch covering lemma on $A$ in \eqref{eq:fA} to get a collection of  $(d-1)$-dimensional cubes, which overlap at most an $\om$-dependent number of times, and on each of those these cubes, the function $f$ from \eqref{eq:fA} does not oscillate too much (depending on $\lambda$). They use each of these $(d-1)$-dimensional cubes to stack multiple $d$-dimensional cuboids of a small $\lambda$-dependent height on top of each other below the graph of $f$, which is part of $\partial\om$. The one on top will in general not be a cuboid, but its upper boundary is given by the corresponding part of the graph of $f$. For this oscillatory domain $D$ on top, Netrusov and Safarov use a new Poincaré-Sobolev inequality \cite[Lemma 2.6.(2)]{netrusov2005weyl} to show that $-\Delta^N_D - \lambda$ has  at most one negative eigenvalue. Netrusov and Safarov can estimate the number of oscillatory demands close to the boundary $\partial\om$ by a constant times $\lambda^{\frac{d-1}{2\ga}}$, so the boundary contribution for $\gamma>\frac{d-1}{d}$ is of subleading order.

\section{Proof strategy of Theorem \ref{th:weightednorm}}
We will explain the proof strategy of the following stronger version of Theorem \ref{th:weightednorm}  for $\gamma\in \left[\tfrac{d-1}{d}, 1\right)$, see \cite[Section 1.2]{dietze2023semiclassical}. Theorem \ref{th:weightednorm} follows from  Theorem \ref{th:weightednormfullrange} combined with \cite[Lemma 5.1]{dietze2023semiclassical}.

\begin{theorem}[Cwikel-Lieb-Rozenblum type bound for $\gamma\in \left[\tfrac{d-1}{d}, 1\right)$]\label{th:weightednormfullrange}
   Let $d\ge 2$. Let $\gamma\in \left[\tfrac{d-1}{d}, 1\right)$ and let $\Omega$ be a $\gamma$-Hölder domain. Then there exists a constant $C_\om = C_\om(d, \gamma, \Omega)> 0$ such that for every $V : \Omega \rightarrow \left(- \infty, 0 \right]$ with $\normiii{V} < \infty$, we have
    \begin{equation}\label{eq:clrthm}
        \ev{\Omega} \leq C_\om\left(1 + \normiii{V}^\frac{d}{2}\right) .
    \end{equation}
    Here the norm $ \normiii{V}=\no{V}{\pt,\be}$ is given in \cite[Definition 2.5]{dietze2023semiclassical}, with $\beta=\beta(d,\ga)>0$ and $\pt=\pt(d,\ga)\in\left(\frac{d}{2},\infty\right)$ chosen as in \cite[equations (35) and (36)]{dietze2023semiclassical}, see also \eqref{eq:triplenormdef} below. 
\end{theorem}

The Cwikel-Lieb-Rozenblum inequality \cite{cwikel1977weak,lieb1976bounds,rozenblum1972} states that for any open set $\om\subset\R^d$ for $d\ge3$ and for every $V : \om \rightarrow \left(- \infty, 0 \right]$, we have
    \begin{equation}\label{eq:clrintbasic2}
        \evd{\om} \leq C(d) \int_\om |V|^{\frac{d}{2}} .
    \end{equation}

    \bigskip

Our proof of Theorem \ref{th:weightednormfullrange} is inspired by both Rozenblum's proof strategy of the Cwikel-Lieb-Rozenblum inequality \eqref{eq:clrintbasic2}, see \cite{rozenblum1971distribution} and \cite[Section 4.5.1]{frank2021schrodinger} and Netrusov's and Safarov's proof strategy of Weyl's law for constant potentials on Hölder domains \eqref{eq:nsintweylnconst} \cite[Corollary 1.6]{netrusov2005weyl}, see Section \ref{s:ns}. 

\bigskip
Since \eqref{eq:clrintbasic2} is an inequality on the entire space  $\R^d$ or on a domain $\om$ with Dirichlet boundary conditions, Rozenblum can choose his oscillatory domains as cubes on the entire domain. In order to make sure that there is at most one negative eigenvalue on each cube, he needs to vary the size of the cubes depending on the size of the potential $V$ there. Using the Besicovitch covering lemma, Rozenblum can make sure that the cubes overlap at most a dimension-dependent number of times, which is sufficient for the Cwikel-Lieb-Rozenblum inequality  since it does not require an optimal constant as opposed to  Weyl's law.

\bigskip

Similarly, for the proof of Theorem \ref{th:weightednormfullrange}, our oscillatory domains need to depend on the size of the potential $V$ locally. Moreover, we have to be careful close to the boundary $\partial\om$, where our oscillatory domains will be given by rectangles intersected with $\om$. Far enough away from the boundary $\partial\om$, these rectangles are cubes.

\bigskip

Close to the boundary, we need to use a new covering lemma for oscillatory domains \cite[Lemma 4.1]{dietze2023semiclassical}. Moreover, in order to relate an oscillatory domain having at most one negative eigenvalue to the size of the potential $V$ on that oscillatory domain measured in an appropriate norm, we use a new Poincaré-Sobolev inequality on oscillatory domains \cite[Corollary 3.3]{dietze2023semiclassical}. Recall that Rozenblum uses the Poincaré-Sobolev inequality on cubes, while Netrusov and Safarov use a Poincaré-Sobolev inequality (only involving the $L^2$ norm) on oscillatory domains \cite[Lemma 2.6.(2)]{netrusov2005weyl}.

\bigskip

Combining all these ingredients, we can bound the number of negative eigenvalues of the Schr\"odinger operator $-\Delta^N_\Omega+V$ by the number of oscillatory domains we chose. In Rozenblum's proof of the Cwikel-Lieb-Rozenblum inequality, the quantity $\int_Q|V|^{\frac{d}{2}}$ was of order one on each cube $Q$. Therefore he could estimate 
\begin{equation}\label{eq:rpfendclr}
\evd{\om} \leq \ \text{number of cubes } Q \ \le C(d) \sum_{\text{cubes  } Q}\int_\om |V|^{\frac{d}{2}}\le C(d) \int_\om |V|^{\frac{d}{2}},
\end{equation}
where he used that the cubes $Q$ can only overlap a finite number of times in the last step.

\bigskip

In our case, we can choose the oscillatory domains $D$ such that $\int_D|V|^{\pt}$ is of order one for some $\pt=\pt(d,\ga)\in\left(\frac{d}{2},\infty\right)$. Imitating the estimate in \eqref{eq:rpfendclr}, we get 
\begin{equation}\label{eq:clrbad}
    N\left(-\Delta^N_\Omega +  V \right) \leq C_{\om}\left(1+\int_{\om}|V|^{\pt}\right),
\end{equation}
where we need to add $1$ on the right-hand side because we are dealing with the Neumann Laplacian, so we can also test with constant functions. However, \eqref{eq:clrbad} is \emph{not} the estimate we aim for. In particular, replacing $V$ by $\lambda V$ in \eqref{eq:clrbad} and letting $\lambda\to\infty$, we obtain 
\begin{equation}\label{eq:clrnicesemibadp}
    \evv{\om}{\lambda V}=\mathcal{O}\left(\lambda^{\pt}\right) \ \textrm{as } \lambda\to\infty,
\end{equation}
which is weaker than \eqref{eq:clrnicesemib} since $\pt>\frac{d}{2}$. The main remaining challenge is to get the expected semiclassical behaviour. The key idea is to use Hölder's inequality for a sum of products of real numbers. If we denote by $\left\{D_j\right\}_{j \in J_3}$ the oscillatory domains that are very close to the boundary, where $J_3$ is an index set, then we can estimate the number  of oscillatory domains very close to the boundary $|J_3|$  for $s, s' \in\left(1, \infty\right)$ with $\tfrac{1}{s} + \tfrac{1}{s'} = 1$, so for any $A_j > 0$, $j \in J_3$, we have
\begin{equation}\label{eq:j3}
    |J_3| = \sum_{j \in J_3} A_j^{-1} A_j \le \left(\sum_{j \in J_3} A_j^{-s'}\right)^{\frac{1}{s'}} \left(\sum_{j \in J_3} A_j^{s}\right)^{\frac{1}{s}} .
\end{equation}
We choose each $A_j > 0$ for $j \in J_3$ such that it only depends on the size of the largest side-length of the oscillatory domain $D_j$ and the distance of its centre to the boundary $\partial\om$ measured in a certain way. The norm $\normiii{V}$ consists of the $L^{\frac{d}{2}}$ norm on $\om$ plus a weighted $L^{\pt}$ semi-norm on $\om$ with a weight $w$ that grows near the boundary at a scale determined by the parameter $\beta=\beta(d,\ga)>0$, that is, on each $D\subset\om$, we have
\begin{equation}\label{eq:triplenormdef}
    \normiii{V}_D:=\left(\inx{D}{x}|V(x)|^{\frac{d}{2}}\right)^{\frac{2}{d}}+\left(\inx{D}{x}w(x)|V(x)|^{\pt}\right)^{\frac{1}{\pt}}.
\end{equation}
The $A_j > 0$ satisfy for all $j \in J_3$ 
\begin{equation}\label{eq:ajs}
    A_j^s \le C \inx{D_j}{x}w(x)|V(x)|^{\pt},    
\end{equation}
so using that the oscillatory domains $\left\{D_j\right\}_{j \in J_3}$ overlap at most a bounded number of times, we get 
\begin{equation}\label{eq:ajssum}
    \sum_{j \in J_3} A_j^s \le C  \inx{\om}{x}w(x)|V(x)|^{\pt}\le C \normiii{V}^{\pt}.
\end{equation}
In fact, we can choose the $A_j > 0$ such that we also have 
\begin{equation}\label{eq:ajmssum}
    \sum_{j \in J_3} A_j^{-s'} \le C \normiii{V}^{-\frac{1}{2}} 
\end{equation}
and we can choose all the parameters such that 
\begin{equation}\label{eq:par}
    \frac{\tilde{p}}{s} - \frac{1}{2s'} = \frac{d}{2} .
\end{equation} 
Combining \eqref{eq:j3}, \eqref{eq:ajssum}, \eqref{eq:ajmssum} and \eqref{eq:par}, we obtain
\begin{equation}\label{eq:j3end}
    |J_3| \le C\normiii{V}^{- \frac{1}{2s'}}\normiii{V}^{\frac{\tilde{p}}{s}}=C\normiii{V}^{\frac{d}{2}},
\end{equation}
which has the desired semiclassical behaviour \eqref{eq:clrnicesemib}.
\printbibliography
\end{document}